\documentstyle[aps]{revtex} 
\begin{document}
\twocolumn
%-----------------------
\title{Hawking radiation without black hole entropy}
%-----------------------
\author{
Matt Visser
}
%----------------------------
\address{
Physics Department, Washington University,
Saint Louis, Missouri 63130-4899
}
%-------------------
\date{2 December 1997; gr-qc/9712016}
%------------------
\maketitle
%-------------------

%------------------------------------------------------------------------------
%\section*{Abstract}
%------------------------------------------------------------------------------

{\small {\underline{\em Abstract:}} In this Letter I point out that
Hawking radiation is a purely kinematic effect that is generic to
Lorentzian geometries. Hawking radiation arises for any test field
on any Lorentzian geometry containing an event horizon regardless
of whether or not the Lorentzian geometry satisfies the dynamical
Einstein equations of general relativity. On the other hand, the
classical laws of black hole mechanics are intrinsically linked to
the Einstein equations of general relativity (or their perturbative
extension into either semiclassical quantum gravity or string-inspired
scenarios).  In particular, the laws of black hole thermodynamics,
and the identification of the entropy of a black hole with its
area, are inextricably linked with the dynamical equations satisfied
by the Lorentzian geometry: entropy is proportional to area (plus
corrections) if and only if the dynamical equations are the Einstein
equations (plus corrections). It is quite possible to have Hawking
radiation occur in physical situations in which the laws of black
hole mechanics do not apply, and in situations in which the notion
of black hole entropy does not even make any sense.  This observation
has important implications for any derivation of black hole entropy
that seeks to deduce black hole entropy from the Hawking radiation.}

%---------------------------------------------------------------------
%\section{Introduction}
%---------------------------------------------------------------------

{\underline{\em Introduction:}} In Einstein gravity (general
relativity), and in theories that perturbatively reduce to Einstein
gravity, the notion of black hole
entropy~\cite{Bekenstein73,Bekenstein74} is inextricably tied up with
the existence of the Hawking radiation
phenomenon~\cite{Hawking74,Hawking75}. Historically the notions were
developed contemporaneously, and served to reinforce one another. The
laws of black hole mechanics were developed first~\cite{Laws}, with
the formal similarity between the second law of black hole mechanics
and the second law of thermodynamics then serving to suggest that
black holes could be assigned an entropy~\cite{Bekenstein73,Laws}. But
it was not until after the discovery of the Hawking radiation
phenomenon~\cite{Hawking74,Hawking75}, that the notion of black hole
entropy became widely accepted, the laws of black hole mechanics then
being promoted to the laws of black hole
thermodynamics~\cite{Bekenstein74,Wald}.

However, with hindsight it is now possible to look back and realise
that these two notions are actually rather distinct in their genesis,
and that there are physical situations (not Einstein gravity) in
which the two notions can be completely divorced --- so that Hawking
radiation can occur even in situations where the very notion of
black hole entropy is meaningless. (For example, as I shall argue
below, the acoustic black holes of Unruh~\cite{Unruh81}, the
solid-state black holes of Reznik~\cite{Reznik97}, and the lattice
black holes of Corley and Jacobson~\cite{Corley-Jacobson97}. See
the discussion in~\cite{Visser97}.)

In this Letter I emphasize that Hawking radiation is a purely
kinematic phenomenon: It occurs in generic Lorentzian geometries
containing event horizons whenever one introduces a test field that
propagates in an (approximate) Lorentz invariant manner. This should
in fact have been realized immediately from the fact that Hawking's
original derivation~\cite{Hawking74,Hawking75} makes no use of the
Einstein equations. However, one would never think to even ask this
question until after the advent of physical models of Lorentzian
geometry that are distinct from Einstein gravity. The fact that sound
waves in a flowing fluid couple to an acoustic metric that defines a
Lorentzian geometry completely unconnected with the propagation of
light is the best known example of such a physical
system~\cite{Unruh81,Visser97,Jacobson91,Comer,Visser93,Jacobson93,%
Unruh94,Hochberg}. A clear pedagogical presentation of the notions
of ergosphere, apparent horizon, event horizon, ``surface gravity''
and ``acoustic black hole'' in the acoustic model of Lorentzian
geometry is presented in~\cite{Visser97}. In fact, it is now known
that the Hawking radiation process is sufficiently robust that
approximate low-energy Lorentz invariance is quite sufficient to
guarantee a thermal spectrum (subject to greybody distortion
factors)~\cite{Reznik97,Corley-Jacobson97,Jacobson91,Jacobson93,Unruh94,%
Brout,Jacobson95,Jacobson96,Corley-Jacobson96,Corley97a,Corley97b,Reznik96}.

On the other hand, black hole entropy, and in fact all of black
hole thermodynamics and the classical laws of black hole mechanics,
are intrinsically dynamical phenomena in that they depend critically
on the perturbative validity of the Einstein equations. This can
be seen from the modern derivations of the various laws of black
hole mechanics~\cite{Laws,Wald}, which proceed by explicitly invoking
the Einstein equations together with the various classical energy
conditions of Einstein gravity~\cite{Wald,Hawking-Ellis,Visser95}. (This
fact is somewhat obscured in some of the early discussions of black
hole mechanics where consideration is implicitly limited to the
standard Schwarzschild, Reissner-Nordstr\"om, Kerr, and Kerr-Newman
black holes.)  That the laws of black hole mechanics generally fail
for the acoustic black holes of the acoustic Lorentzian geometries
is explicitly pointed out in~\cite{Visser97}.

The impact of these results is perhaps a little subtle: The existence
of the Hawking flux in any candidate theory of quantum gravity is
not itself a test of any dynamical aspect of quantum gravity. The
existence of the Hawking flux is not even a test of the dynamics
of the low-energy effective theory. Instead, the Hawking flux tests
the extent to which the candidate theory of quantum gravity is
capable of reproducing the Lorentzian manifold structure that we
have by now come to believe is an inescapable part of the kinematics
of any phenomenologically acceptable theory of gravity (at least
in the low-energy limit probed by current experiments~\cite{Will}).

It is only after one imposes (or derives) dynamical equations for the
low energy effective theory, and only provided that these low-energy
dynamical equations are the Einstein equations (possibly plus
higher-order corrections), that we can invoke the laws of black hole
thermodynamics to see that black holes can be assigned entropies
proportional to their area (possibly plus higher-order
corrections)~\cite{Wald93,Iyer-Wald,Visser93-area,Visser93-surface,%
Jacobson-Kang-Myers93}.  Thus a calculation of the Hawking flux, in
any candidate theory of quantum gravity, supports the notion of black
hole entropy only insofar as it provides reasons for believing the
perturbative applicability of the dynamics encoded in the Einstein
equations.

These conclusions hold independently of whatever model one wishes
to propose for quantum gravity, as they require knowledge only of
the low-energy sub-Planckian phenomenology --- where we at least
think we understand the basic issues.

%---------------------------------------------------------------------
%\section{Acoustic Lorentzian geometries}
%---------------------------------------------------------------------

{\underline{\em Acoustic Lorentzian geometries:}} The
acoustic model for Lorentzian geometry is not widely-known outside of
the confines of the general relativity community so I shall provide a
brief description here.  The model arises from asking the deceptively
simple question of how sound waves propagate in a flowing fluid.
Under suitable restrictions (vorticity-free flow, barotropic equation
of state, zero viscosity) it can be shown that linearizing the
combined Euler and continuity equations of non-relativistic fluid
mechanics leads to sound waves (phonons) that are described by a
scalar field. This phonon is a massless scalar field that is minimally
coupled to the ``acoustic
metric''~\cite{Unruh81,Visser97,Visser93,Unruh94}.  The acoustic
metric is an {\em algebraic} function of the density, speed of sound,
and velocity of the flowing fluid explicitly given by
\begin{equation}
g_{\mu\nu}(t,\vec x) 
\equiv {\rho\over c} 
\left[ \matrix{-(c^2-v^2)&\vdots&-{\vec v}\cr
               \cdots\cdots\cdots\cdots&\cdot&\cdots\cdots\cr
	       -{\vec v}&\vdots& I\cr } \right].
\end{equation}
(Here $I$ is the $3\times3$ identity matrix.) The equation of motion
for the phonon field is simply the usual d'Alembertian
equation~\cite{Unruh81,Visser97,Visser93,Unruh94}
\begin{equation}
\Delta \psi \equiv 
{1\over\sqrt{-g}} 
\partial_\mu 
\left( \sqrt{-g} \; g^{\mu\nu} \; \partial_\nu \psi \right) = 0.
\end{equation}
This model is sufficiently rich to enable probing of almost all of the
kinematic aspects of general relativity (the existence of a Lorentzian
geometry), without the dynamics (the Einstein equations).  The
dynamics of the acoustic Lorentzian geometry are of course governed by
the ordinary nonrelativistic Euler and continuity equations. (The
acoustic Lorentz geometries are not completely arbitrary in that they
automatically satisfy the stable causality
condition~\cite{Visser97,Visser93}, which thereby precludes some of
the more entertaining causality related problems that can arise in
Einstein gravity.) Nevertheless, the acoustic Lorentz geometries are
sufficiently general so as to contain ergospheres, trapped regions,
apparent horizons, event horizons (absolute horizons), and the full
panoply of technical machinery for the kinematic aspects of black hole
physics~\cite{Visser97}. Black holes are defined as regions from which
phonons (which are represented by null geodesics of the acoustic
metric) cannot escape --- because the fluid is flowing inward at
greater than the local speed of sound. At the (future) event horizon
the normal component of the fluid velocity is inward pointing and
equals the local speed of sound, $v_\perp = c$.  The notion of
``surface gravity'' can be defined as for general relativistic black
holes; and for stationary flows measures the extent to which the
natural time parameter defined by the timelike Killing vector fails to
be an affine parameter for those null geodesics that just skim the
event horizon. The ``surface gravity'' can be calculated to
be~\cite{Visser97}
\begin{equation}
g_H = 
{1\over2} \; {\partial(c^2-v_\perp^2)\over \partial n} =
 c \; {\partial(c-v_\perp)\over\partial n}.
\end{equation}
This generalizes the result of Unruh~\cite{Unruh81,Unruh94} to the
case where the speed of sound is position dependent and/or the
acoustic horizon is not the null surface of the time translation
Killing vector. This result is also compatible with that deduced for
the solid-state black holes of Reznik~\cite{Reznik97}, the lattice
black holes of Corley and Jacobson~\cite{Corley-Jacobson97}, and with
the ``dirty black holes'' of~\cite{Visser92}. In the special case
where the speed of sound is independent of position, and the fluid
impinges on the event horizon at right angles, (e.g. if the geometry
is static rather than just stationary) the surface gravity is
identical to the ordinary three-dimensional acceleration of the fluid
as it crosses the horizon~\cite{Visser97}.

As originally discussed by Unruh~\cite{Unruh81}, (and subsequent
papers~\cite{Jacobson91,Jacobson93,Unruh94,Hochberg,Brout,Jacobson95,%
Jacobson96,Corley-Jacobson96,Corley97a,Corley97b,Reznik96})
an acoustic event horizon will emit Hawking radiation in the form of a
thermal bath of phonons at a temperature
\begin{equation}
k \; T_H = {\hbar \; g_H\over 2\pi \; c}.
\end{equation}
(Yes, this really is the speed of sound in the above equation, and
$g_H$ is really normalized to have the dimensions of a physical
acceleration.) This result also applies, with suitable modifications,
to the solid-state black holes of Reznik~\cite{Reznik97} and the
lattice black holes of Corley and Jacobson~\cite{Corley-Jacobson97}.
Using the numerical expression
\begin{equation}
T_H = 
(1.2\times 10^{-9} K m) \;
\left[ {c\over 1000 m s^{-1}} \right]\; 
\left[ {1\over c} {\partial(c-v_\perp)\over\partial n} \right],
\end{equation}
it is clear that experimental verification of this acoustic Hawking
effect will be rather difficult. (Though, as Unruh has pointed
out~\cite{Unruh81}, this is certainly technologically easier than
building [general relativistic] micro-black holes in the laboratory.)

Despite the technological difficulties inherent in bringing these
acoustic black holes to experimental realization, they already provide
us with a clean theoretical laboratory that sharply divorces the
kinematic aspects of general relativity (Lorentzian geometry) from the
dynamic aspects (the Einstein equations). That such a divorce is even
possible in physically realisable systems was not clear before the
advent of the acoustic Lorentzian geometries.

Now that we have at least one clean theoretical laboratory that
makes this separation, theorists can calmly take the next step and
even divorce themselves from the underlying fluid mechanics ---
now turning interest to  Lorentzian geometries in general without
making {\em any} commitment to any particular geometrodynamics, be
it Einstein geometrodynamics or Euler geometrodynamics (acoustic
geometrodynamics). Once this critical conceptual step is made, it
is clear that the calculations
of~\cite{Unruh81,Reznik97,Corley-Jacobson97,Jacobson91,Jacobson93,Unruh94,%
Brout,Jacobson95,Jacobson96,Corley-Jacobson96,Corley97a,Corley97b,Reznik96},
though they were inspired by the acoustic model, actually prove
that Hawking radiation is a completely kinematic effect independent
of {\em any} underlying dynamics for the Lorentzian
geometry~\cite{Visser97}.

%---------------------------------------------------------------------
%\section{Black hole mechanics}
%---------------------------------------------------------------------

{\underline{\em Black hole mechanics:}} The dynamical origin
of the laws of black hole mechanics is evident from the fact that the
various proofs in the literature explicitly use either the Einstein
equations plus the energy conditions~\cite{Laws,Wald,Hawking-Ellis,Visser95},
or at an absolute minimum, the existence of a diffeomorphisim
invariant Lagrangian governing the evolution of the Lorentzian
geometry~\cite{Wald93,Iyer-Wald}. For instance:

\underline{Zeroth law:} In Einstein gravity the constancy of the
surface gravity ({\em mutatis mutandis} the Hawking temperature) over
the event horizon of a stationary black hole (with Killing horizon)
follows from the Einstein equations plus the Dominant Energy
Condition~\cite[pages 331-334]{Wald}. In the acoustic model it is not
even necessary for the event horizon of a stationary acoustic black
hole to be a Killing horizon~\cite{Visser97}, and with sufficiently
complicated fluid flows one can set up arbitrarily complicated
patterns of ``surface gravity''. In general relativity the fact that
stationary but non-static black holes are axisymmetric is deduced from
the fact that non-axisymmetric black holes are expected to loose
energy via gravitational radiation and so dynamically relax to an
axisymmetric configuration --- in the fluid dynamic models there is no
particular reason to even consider dynamically relaxation since the
flow can be maintained by external forces.  With no Einstein
equations, no energy conditions, and not even the guarantee of a
Killing horizon, there is no zeroth law~\cite{Visser97}.

\underline{First law:} The most general derivation of the first law of
black hole mechanics still requires a dynamical evolution for the
Lorentzian geometry that is governed by a diffeomorphisim invariant
Lagrangian~\cite{Wald93,Iyer-Wald}. Subject to this restriction the
entropy of a black hole can be defined in terms of the Lagrangian
governing the dynamics
by~\cite{Wald93,Iyer-Wald,Visser93-area,Visser93-surface,Jacobson-Kang-Myers93}
\begin{equation}
S = k_B \int_H 
{\delta {\cal L} \over \delta R_{\mu\nu\sigma\rho}} \;
\epsilon_{\nu\nu} \epsilon_{\sigma\rho} \;
\sqrt{{}^2 g} \; d^2x.
\end{equation}
The integral runs over some suitable cross-section of the horizon.
If the Lagrangian is Einstein--Hilbert (possibly plus corrections),
then the entropy will be proportional to the area (plus
corrections)~\cite{Wald93,Iyer-Wald,Visser93-area,Visser93-surface,%
Jacobson-Kang-Myers93}.  In the absence of a covariant dynamics,
it does not even make sense to assign an entropy to the event
horizon.

\underline{Second law:} Proofs of the second law of black hole
mechanics, the Hawking area increase theorem, explicitly
invoke the Einstein equations plus the null energy
condition~\cite{Wald,Hawking-Ellis}. Proofs of the generalized second
law of black hole thermodynamics (the increase of total entropy,
ordinary entropy plus black hole entropy) implicitly invoke covariant
dynamics arising from a diffeomorphisim invariant Lagrangian (via
appeal to the first law)~\cite{Frolov-Page,Jacobson-Kang-Myers95,GSL},
and sometimes make even more specific model-dependent assumptions
about the matter fields~\cite{GSL}.

\underline{Third law:} The third law of black hole mechanics (the
impossibility of reaching extremality), is again intrinsically
dynamical. There is considerable ambiguity on how to precisely
formulate the third law (Nernst theorem), and I shall direct
interested readers to the recent paper by Wald~\cite{Wald97}.

%---------------------------------------------------------------------
%\section{Semiclassical quantum gravity}
%---------------------------------------------------------------------

{\underline{\em Semiclassical quantum gravity:}} With this
build up, it is now clear that semiclassical quantum gravity will always
be well behaved with regard to the issues raised in this
Letter. Because semiclassical quantum gravity still models the
universe by a Lorentzian spacetime, there will still be Hawking
radiation from any event horizon. Because semiclassical quantum
gravity has an effective action that is the Einstein-Hilbert term
plus higher-curvature corrections, the black hole entropy will be
proportional to the area plus corrections. Since any putative model
for quantum gravity must reduce to semiclassical quantum gravity in
the sub-Planckian limit, the calculation of emission rates and
entropies in candidate models of quantum gravity are excellent
consistency checks that these models must satisfy.

If one has a complicated model for quantum gravity, then the
complicated calculations required to verify these non-trivial low-energy
consistency checks may indeed seem miraculous --- but the miracle is
more a reflection of the complexity of the candidate model for quantum
gravity than it is any guarantee of the physical correctness of the
model.

%---------------------------------------------------------------------
%\section{String-inspired scenarios}
%---------------------------------------------------------------------

{\underline{\em String-inspired scenarios:}} The quantum gravity
models most ardently being pursued at this stage are the various
string-inspired scenarios.  Recent progress in calculating the
Hawking flux in these string-inspired models, including (in some
limits) greybody factors, and progress in constructing a microphysical
statistical mechanics analog for the black hole entropy has generated
much excitement.  See, for instance,
\cite{Horowitz-Polchinski,Horowitz,Maldacena-Strominger,Cvetic-Larsen,%
Gubser,Emparan}.  Most intriguingly, many of these results seem to
be largely independent of the technical details of fundamental
string theory~\cite{Maldacena-Strominger}.  From the point of view
argued in this Letter, this is only natural: By showing that in
certain limits string-inspired models produce a Hawking flux one
is verifying that in these limits the string models are compatible
with the existence of a Lorentzian geometry with event horizon.
By calculating the precise spectra of the string-model Hawking
fluxes (greybody factors), and comparing them with the greybody
factors for the canonical black holes (Schwarzschild, Reissner-Nordstr\"om,
Kerr, Kerr-Newman) one is testing the extent to which string-inspired
models reproduce standard physics.  In fact, calculating greybody
factors is a complicated way of implicitly checking that the
low-energy dynamics is in fact Einstein-Hilbert.  (A warning: since
we expect the low-energy limit of string models to reduce to Einstein
gravity plus stringy corrections, we should expect stringy black
holes to be canonical black holes plus stringy corrections, and so
we should not expect the greybody factors to be identical to the
canonical ones beyond lowest nontrivial order in the string tension.)

Thus I argue that calculations of greybody factors in string-inspired
models for quantum gravity are best viewed as non-trivial consistency
checks on the ability of these models to be compatible with a
suitable low-energy semiclassical quantum gravity. These consistency
checks are not unique to the string-inspired models and must be
faced by {\em any} candidate for quantum gravity. The feature of
the string-inspired models that is more important than the precise
form of the greybody factors is the fact that these scenarios appear
to provide a unitary description of the Hawking flux.

Turning to the black hole entropy: Since we believe that the
low-energy limit of the string models are point-particle field
theories defined on Lorentzian geometries, and that in this limit
the string models reproduce Einstein gravity, then we automatically
know (without further calculation) that black holes can be assigned
an entropy proportional to their area. The feature of the
string-inspired models that goes beyond semiclassical quantum
gravity is the fact that this notion of entropy can be continuously
extended to regions of parameter space where black holes do not
exist and explicit state counting calculations hold sway.

%----------------------------------------------------------------------------
%\section*{Acknowledgements}
%----------------------------------------------------------------------------

\underline{\em Acknowledgements:}
I wish to thank Ted Jacobson for helpful comments. 
This research was supported by the US Department of Energy. 
%----------------------------------------------------------------------------

%----------------------------------------------------------------------------
\end{document}